\documentclass[prb,preprint]{revtex4-1} 
\usepackage{amsmath}  % needed for \tfrac, \bmatrix, etc.
\usepackage{amsfonts} % needed for bold Greek, Fraktur, and blackboard bold
\usepackage{graphicx} % needed for figures
\usepackage{bm}
\usepackage{float}
\usepackage{hyperref}
\hypersetup{pdfborderstyle={/S/U/W 1}}

\usepackage{subfig} % For sub-panels to figures
\usepackage[countmax]{subfloat}

\begin{document}

% Be sure to use the \title, \author, \affiliation, and \abstract macros
% to format your title page.  Don't use lower-level macros to  manually
% adjust the fonts and centering.

\title{The mobile phone as a free-rotation laboratory}
% In a long title you can use \\ to force a line break at a certain location.

\author{Michael S.~Wheatland}
\email{michael.wheatland@sydney.edu.au} % optional
\author{Tara Murphy}
\author{Daniel Naoumenko}
\author{Daan van Schijndel}
\author{Georgio Katsifis}
\affiliation{School of Physics, The University of Sydney, NSW 2006, Australia}

% See the REVTeX documentation for more examples of author and affiliation lists.

\date{\today}

\begin{abstract}
Modern mobile phones contain a three-axis microelectromechanical system (MEMS) gyroscope, capable of taking accurate measurements of the angular velocity along the three principal axes of the phone with a sampling rate of 100~Hz or better. If the phone is tossed in the air, then, neglecting air resistance, it is in free rotation (rotation in the absence of a torque) with respect to its centre of mass, and the phone's gyroscope can be used to record the rotational dynamics. This enables experimental investigation of free rotation. In this paper, we use a mobile phone to demonstrate the steady states for rotation of the phone about two of its principal axes, and the instability in rotation about the third corresponding to the intermediate moment of inertia. We also show the approximate conservation of angular momentum and rotational kinetic energy during motion in the air, and compare the data with numerical solution of Euler's equations for free rotation. Our results demonstrate the capability of smartphones for investigating free rotation, and should be of interest to college and university teachers developing ``at home'' physics labs for remote learning.
\end{abstract}
% AJP requires an abstract for all regular article submissions.
% Abstracts are optional for submissions to the "Notes and Discussions" section.

\maketitle % title page is now complete

\section{Introduction} % Section titles are automatically converted to all-caps.
% Section numbering is automatic.

Modern mobile phones (smartphones) are versatile data collection devices for experimental physics.\cite{Kuhn-Vogt-2015,Hochberg-etal-2018} Mobile devices such as smartphones and tablets can measure accelerations, rotations, magnetic fields, pressure, and sound and light levels, and are also able to record and analyze video.\cite{Becker-etal-2016,Becker-etal-2020a,Becker-etal-2020b,Hochberg-etal-2020} They are increasingly used in undergraduate physics laboratories, and they can easily facilitate ``at-home'' labs because, in many countries, most students will own a mobile phone. The use of smartphones for this purpose increased in 2020 with the need to rapidly develop online undergraduate physics labs during the COVID-19 crisis.\cite{Bobroff-et-al-2020,Wright-2020} 

Modern smartphones measure acceleration and angular velocity along three axes that are fixed with respect to the phone using microelectromechanical system (MEMS) inertial sensors. The three-axis MEMS gyroscope is a relatively recent addition to phones: for example it first appeared in the iPhone in 2010.\cite{ifixit-2010} In applications designed to use the sensor, the gyroscope sampling rate is typically 100~Hz or 200~Hz,\cite{Michalevsky-2014} which allows data collection during fairly rapid movements of the phone. The accelerometer and gyroscope can be used for physics experiments, including measuring the acceleration due to gravity,\cite{Vogt-Kuhn-2012a,Berrada-etal-2020a,Berrada-etal-2020b} measuring moments of inertia,\cite{Patrinopoulos-Kefalis-2015,Kaps-Stallmach-2020} demonstrating the parallel axis theorem,\cite{Salinas-etal-2019} exploring pendulum dynamics,\cite{Vogt-Kuhn-2012b} and measuring the radial acceleration and Coriolis force in circular motion.\cite{Vogt-Kuhn-2013,Hochberg-et-al-2014,Shakur-Craft-2016}

If a mobile phone is tossed in the air, then, neglecting air resistance, its centre of mass is in simple projectile motion, and the phone is in free rotation (rotation in the absence of a torque) about the centre of mass. Free rotation provides a beautifully simple demonstration of rotational dynamics. It also exhibits counter-intuitive results, including the intermediate axis theorem: for an object with three different moments of inertia about its principal axes, like a mobile phone, the axis with the intermediate moment of inertia is unstable to rotation.\cite{Landau-Lifshitz-1969,Goldstein-1980} As a consequence, if the object is initially set rotating only about the intermediate axis, it will start to rotate about the other two axes. 
Moreover, with the right initial conditions the object can execute a 180$^\circ$ turn about the axis with the smallest moment of inertia, while performing a 360$^\circ$ turn about the intermediate axis. This is called the ``tennis racquet effect,'' or the Dzhanibekov effect.\cite{Murakami-etal-2016,Van-Damme-etal-2017,Mardesic-etal-2020} It can be demonstrated by tossing a tennis racquet upwards from an initially horizontal position: after the racquet rotates about the unstable axis by 360$^\circ$ the opposite face is upwards, showing that it has also made a half-turn about its long axis. A more extreme example is provided by spinning wingnuts and t-handles in zero gravity.\cite{t-handle} Their orientation in space repeatedly reverses when spinning rapidly about the unstable axis. This striking consequence of the instability was identified by Dzhanibekov.\cite{Trivailo-Kojima-2019}

It is easy to demonstrate the tennis racquet effect with a mobile phone, with the MEMS gyroscope collecting data on the rotational dynamics. The instability in rotation about the intermediate axis was previously demonstrated using phone sensors, but the investigation was qualitative, and presented only linear acceleration data.\cite{Loth-etal-2017}

In this paper, we analyse gyroscopic data from a mobile phone tossed in the air and show how it enables a variety of basic physics experiments, including demonstration of the approximate conservation of rotational kinetic energy and the magnitude of angular momentum, as well as a quantitative study of the tennis racquet effect. To collect data we use the app phyphox\cite{phyphox} (RWTH Aachen University), which is available for iOS and Android phones. The data presented here are from an iPhone 8 Plus, but the results have been reproduced on a variety of other phones. 

% Other examples? Interesting questions?

% Summarise content of paper.

\section{Theory}\label{sec:theory}

\begin{figure}[H]
\centering
\includegraphics[width=4cm]{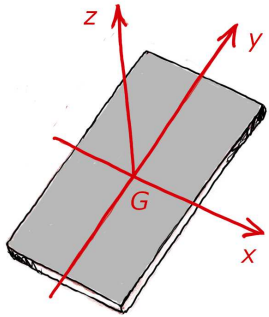}
\caption{The three principal axes ($x$, $y$, and $z$) of the phone. The centre of mass is $G$.}
\label{principal-axes}
\end{figure}

We consider the motion of the phone using the principal-axis coordinate system $Gxyz$ shown in Fig.~\ref{principal-axes}. The origin $G$ is the centre of mass of the phone. The $y$-axis is chosen such that the positive direction is from the bottom of the phone towards the top, and the $z$-axis is directed outwards from the front face of the phone. 

The rotational equation of motion is 
\begin{equation}\label{eq:euler1}
\left.\frac{{\rm d}{\bf L}}{{\rm d}t}\right|_{Gxyz}+\bm{\omega}\times {\bf L}=\bm{\tau},
\end{equation}
where ${\bf L}$ is the angular momentum, $\left.\frac{{\rm d}{\bf L}}{{\rm d}t}\right|_{Gxyz}$ is the rate of change of the angular momentum in the rotating frame $Gxyz$, $\bm{\omega}$ is the angular velocity of the phone (which is the rate of rotation of $Gxyz$ in the Newtonian reference frame), and $\bm{\tau}$ is the net external torque about the centre of mass. The left-hand side of Eq.~(\ref{eq:euler1}) separates the rate of change of the angular momentum into the rate of change with respect to $Gxyz$, and the change due to the rotation of the frame. This approach can be used to calculate the rate of change of any vector in the rotating frame.\cite{Landau-Lifshitz-1969,Goldstein-1980}
The angular velocity is
\begin{equation}
\bm{\omega}=\omega_x\widehat{\bm{x}}+\omega_y\widehat{\bm{y}}+\omega_z\widehat{\bm{z}},
\end{equation}
where $\widehat{\bm{x}}$, $\widehat{\bm{y}}$ and $\widehat{\bm{z}}$ are the unit vectors in the body-fixed reference frame $Gxyz$. Because $x$, $y$ and $z$ are the principal axes, the angular momentum is
\begin{equation}
\bm{L}=I_x\omega_x\widehat{\bm{x}}+I_y\omega_y\widehat{\bm{y}}+I_z\omega_z\widehat{\bm{z}},
\end{equation}
where $I_x$, $I_y$, and $I_z$ are the principal moments of inertia.

When the phone is tossed in the air the torque about the centre of mass due to gravity is zero. Air resistance can produce a net torque, but it is usually negligible, so that Eq.~(\ref{eq:euler1}) reduces to Euler's equations:
\begin{eqnarray}\label{eq:euler}
I_x\dot{\omega}_x&=&-(I_z-I_y)\omega_y\omega_z \label{eq:euler2-x}\\
I_y\dot{\omega}_y&=&-(I_x-I_z)\omega_x\omega_z \label{eq:euler2-y}\\
I_z\dot{\omega}_z&=&-(I_y-I_x)\omega_x\omega_y,\label{eq:euler2-z}
\end{eqnarray}
where the dot denotes the time derivative.

The rotational kinetic energy and angular momentum are constants of the motion, given by
\begin{equation}\label{eq:K}
K=\frac{1}{2}I_x\omega_x^2+\frac{1}{2}I_y\omega_y^2+\frac{1}{2}I_z\omega_z^2
\end{equation}
and
\begin{equation}
\bm{L}=I_x\omega_x\widehat{\bm{x}}+I_y\omega_y\widehat{\bm{y}}+I_z\omega_z\widehat{\bm{z}},
\end{equation}
respectively. The components of $\bm{L}$ in the $Gxyz$ frame are not constant, because the frame is 
rotating. However, it is easy to construct the magnitude of the angular momentum vector from the components:
\begin{equation}\label{eq:Lmag}
L=\sqrt{\left(I_x\omega_x\right)^2+\left(I_y\omega_y\right)^2+\left(I_z\omega_z\right)^2}.
\end{equation}

Mobile phones are approximately rectangular cuboids, so the moments of inertia may be approximated by
\begin{equation}\label{eq:Ix}
I_x=\frac{1}{12}m\left(h^2+d^2\right),
\end{equation}
\begin{equation}\label{eq:Iy}
I_y=\frac{1}{12}m\left(w^2+d^2\right),
\end{equation}
and
\begin{equation}\label{eq:Iz}
I_z=\frac{1}{12}m\left(h^2+w^2\right),
\end{equation}
where $m$, $h$, $w$ and $d$ are the mass, height, width, and depth of the phone, respectively. For the standard phone shape $h>w>d$, and hence $I_y<I_x<I_z$. The $x$-axis is the ``intermediate axis,'' i.e.\ has the intermediate moment of inertia.

It is interesting to consider the steady states of Eqs.~(\ref{eq:euler2-x})-(\ref{eq:euler2-z}). Setting the derivatives to zero gives (assuming distinct moments of inertia):
\begin{equation}
\omega_y\omega_z=\omega_x\omega_z=\omega_x\omega_y=0.
\end{equation}
This implies that, for a steady state, two of the angular velocity components must be zero, i.e.\ a steady state rotation can involve only one principal axis. The possible steady states therefore are 
\begin{equation}\label{eq:steady-states}
\bm{\omega}_1=\pm(L/I_x,0,0),\quad
\bm{\omega}_2=\pm(0,L/I_y,0),\quad
\bm{\omega}_3=\pm(0,0,L/I_z).
\end{equation}

However, only $\bm{\omega}_2$ and $\bm{\omega}_3$ are stable steady states -- the intermediate axis is unstable to steady rotation. To see this, consider the state\cite{Acheson-1997}
\begin{equation}\label{eq:instability}
\bm{\omega}=\bm{\omega}_1+(\varepsilon_x,\varepsilon_y,\varepsilon_z),
\end{equation}
where $\varepsilon_x$, $\varepsilon_y$ and $\varepsilon_z$ are time-dependent, with $|\varepsilon_i|\ll L/I_x$ for $i=x,y,z$. Substituting Eqs.~(\ref{eq:instability}) into Eqs.~(\ref{eq:euler2-x})-(\ref{eq:euler2-z}) and ignoring the quadratic terms in $\varepsilon_i$ gives
\begin{eqnarray}
I_x\dot{\varepsilon}_x&=&0\label{eq:euler-x2}\\
I_y\dot{\varepsilon}_y&=&\pm \left(I_z-I_x\right)\omega_1\varepsilon_z\label{eq:euler-y2}\\
I_z\dot{\varepsilon}_z&=&\pm \left(I_x-I_y\right)\omega_1\varepsilon_y,\label{eq:euler-z2}
\end{eqnarray}
where $\omega_1=L/I_x$.
Equation~(\ref{eq:euler-x2}) gives $\varepsilon_x=\mathrm{const}$, and Eqs.~(\ref{eq:euler-y2}) and~(\ref{eq:euler-z2}) imply
\begin{equation}\label{eq:euler-y3-z3}
\ddot{\varepsilon}_y/\varepsilon_y=\ddot{\varepsilon}_z/\varepsilon_z=\lambda^2
\end{equation}
where
\begin{equation}
\lambda^2=\frac{(I_z-I_x)(I_x-I_y)}{I_yI_z}\omega_1^2
\end{equation}
is positive. Equations~(\ref{eq:euler-y3-z3}) have the solutions $\varepsilon_y=\varepsilon_{y0}{\rm e}^{|\lambda|t}$ and $\varepsilon_z=\varepsilon_{z0}{\rm e}^{|\lambda|t}$. If the spin is initially predominantly about the $x$-axis, any small perturbations in spin about the other two axes grow exponentially, which is the definition of instability. It is straightforward to show using the same approach that the $y$- and $z$-axes are stable for rotation, i.e.\ in these cases perturbations from the steady states lead to oscillations rather than exponential growth.

The linear analysis given here describes the onset of the instability in rotation about the $x$-axis. More complete analyses show how, for a range of initial conditions close to the unstable steady state, one complete rotation about the $x$-axis is accompanied by a nearly-half rotation about the $y$-axis -- the tennis racquet effect.\cite{Murakami-etal-2016,Van-Damme-etal-2017,Mardesic-etal-2020} 

\section{Results}\label{sec:results}

\begin{figure}[ht]
\centering
\includegraphics[width=10cm]{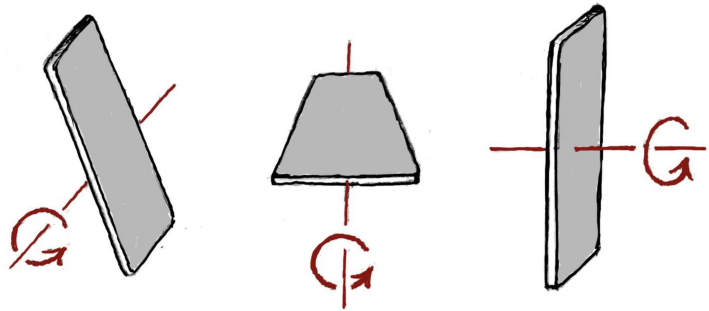}
\caption{Rotations about the three principal axes (the $x$-, $y$-, and $z$-axes, from left to right). We consider experiments involving tossing the phone in the air to achieve rotations predominantly about these axes.}
\label{rotations}
\end{figure}

We consider three experiments in which the mobile phone is tossed in the air, and then caught again. In each case the aim is to set the phone spinning predominantly about one principal axis. The necessary rotations are illustrated in Fig.~\ref{rotations}. The experiments should be performed over a soft surface, in the event of a failed catch!

\subsection{Spinning the phone about the $z$-axis}

\begin{figure}[ht]
\centering
\subfloat[]{
\includegraphics[width=9cm]{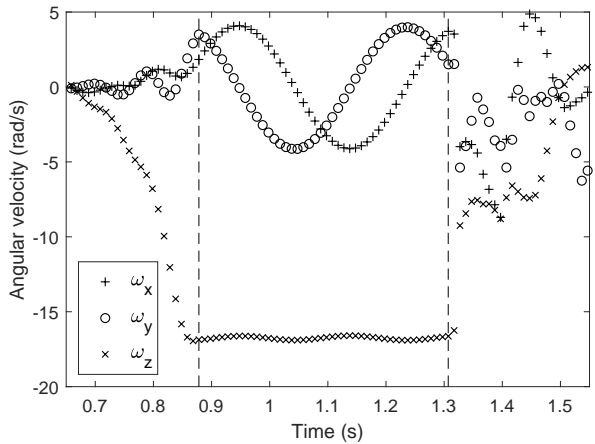}}

\subfloat[]{
\includegraphics[width=9cm]{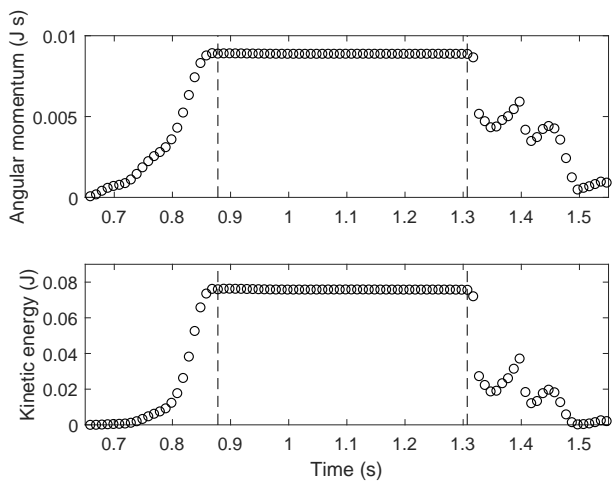}}

\caption{Data collected for a toss (and catch) of a mobile phone producing a spin predominantly about the $z$-axis. The dashed vertical lines indicate the approximate interval during which the phone was in the air. (a) The components of angular velocity about the three principal axes ($x$, $y$, and $z$) versus time. (b) Magnitude of angular momentum and rotational kinetic energy versus time. The variation in $L$ and $K$ is less than 1\% during the time the phone is in the air.\label{zspin}}
\end{figure}

We start with the $z$-axis, which has the largest moment of inertia. An example of data collected using phyphox on the iPhone 8 Plus for a spin about the $z$-axis is shown in Fig.~\ref{zspin}. Panel (a) shows the three angular velocity components in the principal axis frame ($\omega_x$, $\omega_y$, and $\omega_z$) versus time. It is easy to identify the time interval when the phone is in the air (between the dashed vertical lines) by the smooth variation in the angular velocities. The sampling frequency of the gyroscope data ($100$\,Hz) is more than sufficient to resolve the variations in the angular velocities. The angular velocity about the $z$-axis is large and approximately constant ($\omega_z\approx -17$\,rad/s), and the spins about the other two axes are substantially smaller and oscillatory. This is consistent with the $z$-axis being stable to rotation.

It is straightforward to calculate the rotational kinetic energy and the magnitude of the angular momentum of the phone, from the angular velocity data, using Eq.~(\ref{eq:K}) and Eqs.~(\ref{eq:Lmag})-(\ref{eq:Iz}). Table~I lists the parameters of the phone (obtained from the manufacturer's web site\cite{Apple-iPhone-8Plus}) together with the calculated moments of inertia about the principal axes. The uncertainties in $I_x$, $I_y$, and $I_z$ are calculated by assuming an uncertainty of one in the final digit of the stated values of $m$, $h$, $w$, and $d$. 
Panel (b) of Fig.~\ref{zspin} shows the calculated angular momentum magnitude and rotational kinetic energy versus time for our first experiment. The values of $L$ and $K$ are very nearly constant over the interval when the phone is in the air. There are small decreases in $L$ and $K$, which are less than 0.4\% and 0.8\% respectively. This figure provides a remarkable demonstration of the near-conservation of angular momentum and rotational kinetic energy during the motion.

\begin{table}[ht]
\centering
\caption{Properties of the iPhone 8 Plus.}
\begin{ruledtabular}
\begin{tabular}{c c c c c c c}
$m$ (kg) & $h$ (m) & $w$ (m) & $d$ (m) & $I_x$ (kg\,m$^2$) & $I_y$ (kg\,m$^2$) & $I_z$ (kg\,m$^2$) \\
\hline
0.202 & 0.1584 & 0.0781 & 0.0075 & $(4.23\pm 0.02)\times 10^{-4}$ & $(1.04\pm 0.02)\times 10^{-4}$ & $(5.25\pm 0.02)\times 10^{-4}$ \\
\end{tabular}
\end{ruledtabular}
\label{bosons}
\end{table}

\subsection{Spinning the phone about the $y$-axis}

\begin{figure}[ht]
\centering
\subfloat[]{
\includegraphics[width=9cm]{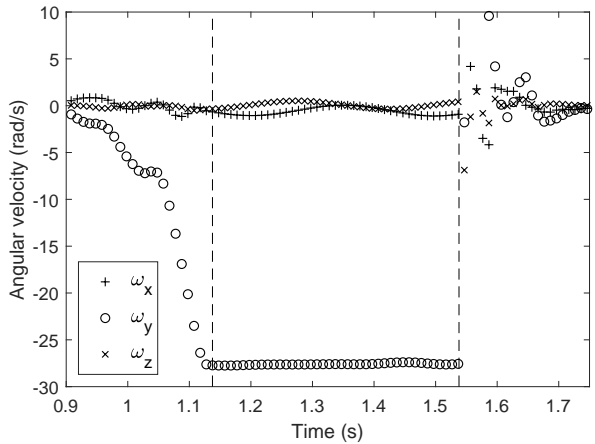}}

\subfloat[]{
\includegraphics[width=9cm]{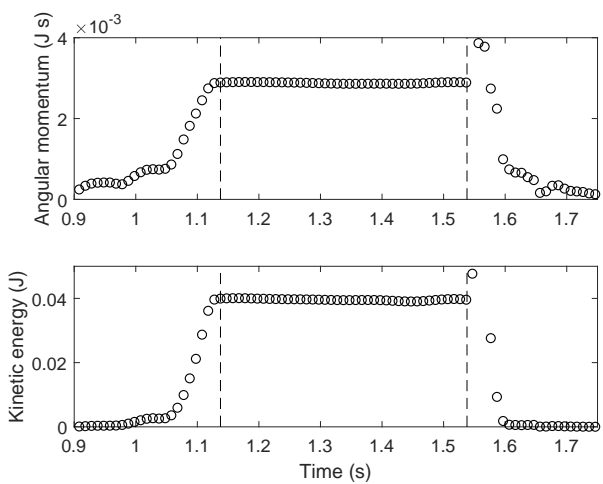}}
\caption{Data for a toss (and catch) of the phone producing a spin predominantly about the $y$-axis. (a) The components of angular velocity about the three principal axes ($x$, $y$, and $z$) versus time. (b) Magnitude of angular momentum and rotational kinetic energy versus time. The variations in $L$ and $K$ are only a few percent during the time the phone is in the air.\label{yspin}}
\end{figure}

Next, we consider the $y$-axis, which has the smallest moment of inertia. Fig.~\ref{yspin} shows the results of an attempt to set the phone spinning only about the $y$-axis. This requires giving the phone a flick about its long axis, as it is tossed, and then caught. Panel (a) shows the angular velocity values. The results are similar to Fig.~\ref{zspin}. The dashed vertical lines again identify the approximate interval the phone is in the air. During this time the angular velocity about the $y$-axis is $\omega_y\approx -28$\,rad/s. Because $I_y$ is the smallest of the three moments of inertia, it is easy to achieve a relatively large spin rate around this axis. There are small oscillatory components of the angular velocity about the other two principal axes, which is consistent with the stability of the $y$-axis to rotation. Panel (b) shows the time variation of $L$ and $K$. Once again these quantities are nearly constant during the time in the air. Small variations in $L$ and $K$ are observed, of order 1.5\% and 3\% respectively.

\subsection{Spinning the phone about the $x$-axis}

\begin{figure}[ht]
\centering
\subfloat[]{
\includegraphics[width=9cm]{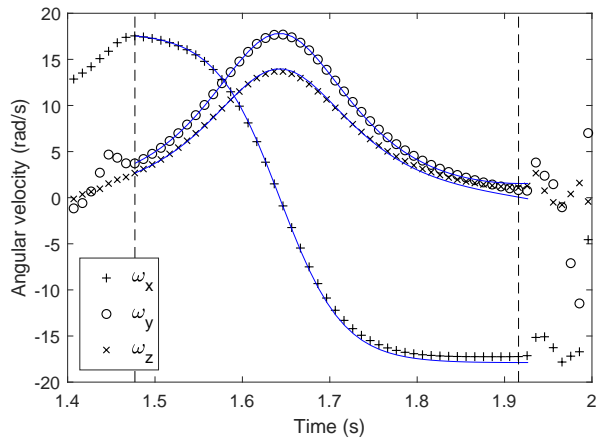}}

\subfloat[]{
\includegraphics[width=9cm]{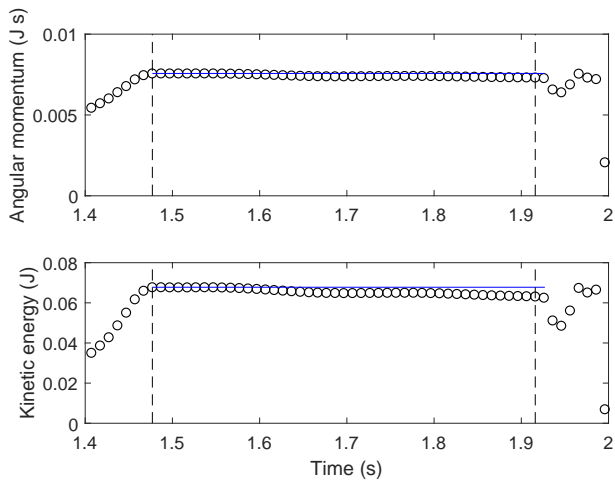}}
\caption{Data for a toss (and catch) of the phone attempting to produce a spin predominantly about the $x$-axis. The numerical solution of the rotational equations of motion is shown as solid blue curves. (a) The components of angular velocity about the three principal axes ($x$, $y$, and $z$) versus time. (b) Magnitude of angular momentum and rotational kinetic energy versus time. The values of $L$ and $K$ decrease by about 3\% and 7\% respectively during the time the phone is in the air. \label{xspin}}
\end{figure}

Finally, we consider the $x$-axis. Fig.~\ref{xspin} shows a typical result of an attempt to toss the phone in the air so that it spins only about the $x$-axis. The results in panel (a), for the angular velocities versus time, are strikingly dissimilar to those in Figs.~\ref{zspin} and~\ref{yspin}. It is again easy to identify the interval of time during which the phone is in the air (indicated by the dashed vertical lines) because of the smooth variation in the angular velocities. The phone initially has a substantial  positive angular velocity about the $x$-axis ($\omega_x\approx 17$\,rad/s), and small components about the other two axes ($\omega_y\approx 4$\,rad/s and $\omega_z\approx 3$\,rad/s), but this situation quickly changes. The components of spin along $y$ and $z$ initially grow exponentially, whilst the rate of change of $\omega_x$ is initially stationary -- results consistent with the linear analysis given in Section~\ref{sec:theory}. The nonlinear development sees the phone reverse its spin about the $x$-axis, so that its final angular velocity is $\omega_x\approx -17$\,rad/s, and the other two components of angular velocity have again fallen to relatively small values. This is the tennis racquet effect. The phone has rotated by 360$^\circ$ around $x$ and rotated by 180$^\circ$ about $y$, leading to a reversal (a flip) of the sign of its spin about $x$. The difference in the results for the $x$-axis compared with $z$ and $y$ will be obvious to a student: the phone refuses to simply flip about the $x$-axis -- it pivots in mid-air around the other axes. The data collected by the gyroscope provides a quantitative confirmation. Panel (b) shows the time variation of $L$ and $K$. In contrast to panel (a), this is quite similar to the results in Figs.~\ref{zspin} and~\ref{yspin}. The angular momentum magnitude and kinetic energy are nearly conserved during the time in the air. A small decrease in each quantity is observed, however, with the angular momentum decreasing by about 3\%, and kinetic energy decreasing by almost 7\%.

We can compare the results with the direct numerical solution of Eqs.~(\ref{eq:euler}) using fourth order Runge-Kutta.\cite{Press-etal-92}. The solid blue curves in Fig.~\ref{xspin} are obtained by integrating the Euler equations, using the initial values of the angular velocity components (the values at the left vertical dashed lines in panel (a)), together with the calculated moments of inertia. Panel (a) shows that the model reproduces the observed time variation in the angular velocities, although a small discrepancy is apparent over the course of the integration. Panel (b) compares the constant model values of $L$ and $K$ with the observations. As noted above there is a small decrease in measured values.

\subsection{Air resistance}

\begin{figure}[ht]
\centering
\includegraphics[width=7cm]{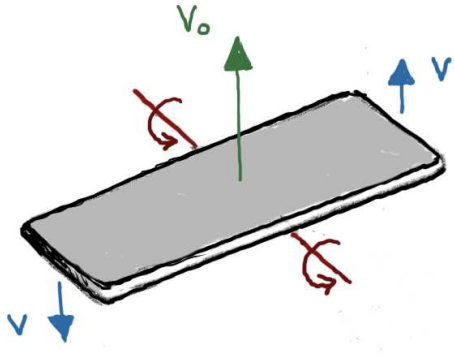}
\caption{Air resistance can produce a torque with respect to the centre of mass. The side of the phone that has an instantaneous velocity in the same direction as the centre of mass motion has a greater relative velocity with respect to the air, and hence experiences a larger drag force.\label{drag-torque}}
\end{figure}

Although gravity cannot produce a torque about the centre of mass, air resistance can. Fig.~\ref{drag-torque} illustrates this point. We assume that at a certain time the centre of mass of the phone is moving upwards through the air with speed $v_0$, and the phone is rotating about the $x$-axis which is perpendicular to the direction of motion. The right side of the phone has a speed $v_0+v$ with respect to the air, and the left side has speed $v_0-v$. Since the drag force is proportional to the square of the relative velocity, the right side experiences a greater drag force, leading to a torque opposing the rotation. Our argument assumes the particular configuration shown, but the result is general: if the centre of mass is moving perpendicular to the rotation axis, air resistance will generate a torque opposing rotation. 

We can estimate the effect of air resistance on the results of our experiments as follows. If we consider the configuration shown in Fig.~\ref{drag-torque}, then the magnitude of the torque is
\begin{eqnarray}
\tau&\approx& \frac{1}{2}\rho\, C_{\rm D}A\ell \left[(v_0+v)^2-(v_0-v)^2\right]\\
&=&2\rho C_{\rm D} A\ell vv_0,
\end{eqnarray}
where $\rho\approx 1.2$\,kg/m$^3$ is the density of air, $C_D$ is the dimensionless drag coefficient, $A\approx \frac{1}{2}hw$ is the area one side of the phone presents to the air, and $\ell\approx h/4$ is the distance at which the force is applied. For our purposes it is sufficient to assume a constant value $v_0\approx \frac{1}{2}gT$ from the time of flight ($T\approx 0.5$\,s), with $g=9.8$\,m/s$^2$, and we can assume $v=\omega \ell$, where $\omega\approx 10$\,rad/s is a typical angular velocity. We will assume $C_D\approx 1$ since measured drag coefficients for non-streamlined objects in air are typically of order unity.\cite{NASA-Drag} Assuming $\Delta L/T\approx \tau$ leads to
\begin{eqnarray}
\Delta L&=&\frac{1}{32}\rho h^3w\omega g T^2\\
&\approx& 3\times 10^{-4}\,\rm{J}\,{\rm s}.
\end{eqnarray}
If we compare this with the total angular momentum $L\approx 7.5\times 10^{-3}$\,{\rm J}\,{\rm s} in the third experiment, then $\Delta L/L\approx 4$\%. This is comparable to the observed decrease in the angular momentum magnitude in this case. 

We can also consider the first experiment, where the rotation was predominantly about the $z$-axis. In this case the area presented to the air was about ten times less, since $d/w\approx 0.1$. The observed change in angular momentum was also about ten times smaller.

These estimates are imprecise but they suggest that air resistance may account for some (or most) of the observed variation in angular momentum and kinetic energy, and hence for the departure between the observations and the numerical solution to Euler's equations. 

There are additional sources of error in these experiments. If the mobile phone is spun too fast, the angular velocity measurements saturate at a ``full-scale'' value.\cite{ST-documentation} For the iPhone 8 Plus the full-scale value is 2000 deg/s, which is about 35 rad/s. This can be demonstrated by tossing the phone so that it is rapidly spinning about the $y$-axis. The MEMS gyroscopes also exhibit nonlinearity, which can be as large as 5\% of the full-scale value. Given the magnitude of the nonlinearity, in principle this effect could also be contributing to the observed departures from the free rotation model. Finally, if we take the uncertainties in Table~I and propagate them through to the values of $L$ and $K$ calculated from the data for the third experiment, the implied errors are about 0.5\% in each case. Hence the error in modeling the moments of inertia is likely to be less important than the other effects.  

\section{Conclusion}\label{sec:conclusion}

The 3-axis MEMS gyroscope in a mobile phone enables a variety of experiments in rotational dynamics. In this paper, we use a phone to investigate free rotation, by tossing and catching the phone while it is collecting data. The stability of rotation about two principal axes, and the instability about the third, intermediate axis, is easy to demonstrate. It is also straightforward to calculate the rotational kinetic energy and angular momentum magnitude from the data, and to show that these quantities are nearly conserved during the time that the phone is in the air.  

Various extensions of the experiments described here are possible. The accelerometer and the gyroscope on the phone could be used to collect data simultaneously, and this could be used to determine the centre of mass motion as well as the rotational motion of the phone. Objects could be attached to the phone, to change the moments of inertia or the principal axes of the combined object. If a calculator is attached to the phone with rubber bands, such that the long axis of the calculator is perpendicular to the long axis of the phone, it is possible to stabilize the rotation about the $x$-axis of the phone. The phone could also be strapped to a half-filled plastic water bottle, which, when tossed, changes its moments of inertia in mid-air, and hence exhibits interesting rotational dynamics.\cite{Dekker-etal-2018} 

\begin{figure}[ht]
\centering
\includegraphics[width=12cm]{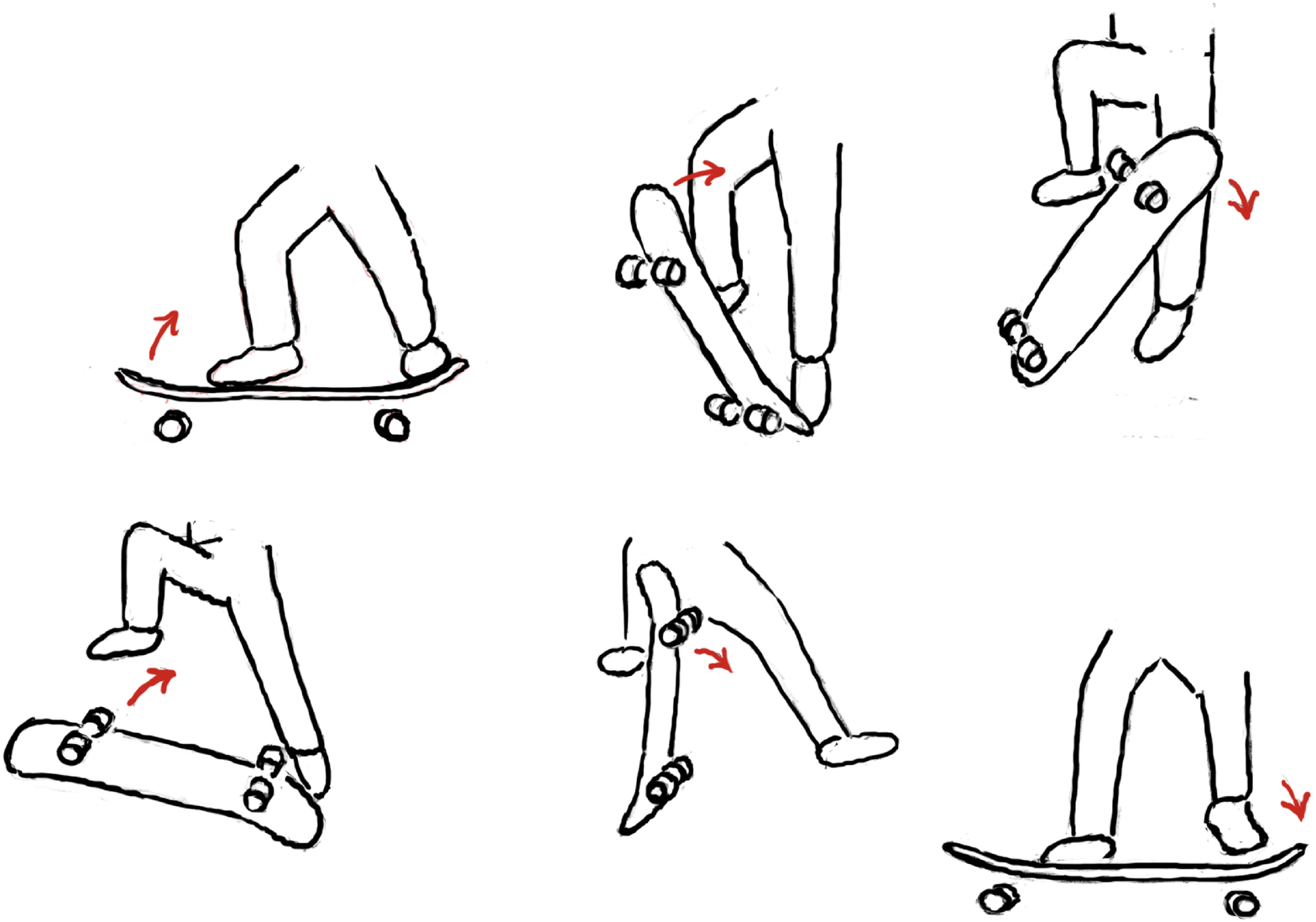}
\caption{The monster flip: a skateboard trick in which the board executes a 360-degree flip about the unstable axis.\label{skateboard}}
\end{figure}

The instability in rotation about the intermediate axis of a three-dimensional rigid body was identified by Poinsot in the 1800s and is discussed in classical mechanics texts,\cite{Landau-Lifshitz-1969,Goldstein-1980} but it has attracted renewed attention with investigations of the tennis racquet and Dzhanibekov effects\cite{Murakami-etal-2016,Van-Damme-etal-2017,Mardesic-etal-2020} and is the subject of recent popular expositions\cite{Veritasium-2019}. The instability and related effects may have applications in spacecraft altitude orientation and quantum control.\cite{Trivailo-Kojima-2019,Mardesic-etal-2020} It is also relevant for skateboard tricks. Different versions exist of the trick of jumping with the skateboard and flipping the board around the unstable axis, and then landing on the board again with the wheels down. One version, called the ``impossible,'' is achieved by guiding the rotation of the board about the unstable axis with the foot.\cite{Physics-Girl-2018} However, a version with the board in free rotation (the ``monster flip'') can also be executed,\cite{Monster-flip-2015} as illustrated in Fig.~\ref{skateboard}. The difficulty of performing this trick whilst avoiding the board exhibiting the tennis racquet effect (so that the board lands wheels-up) has been assessed.\cite{Mardesic-etal-2020}

Experiments on free rotation with a mobile phone are likely to appeal to undergraduate students, and to encourage interest in dynamics. The material here is a suitable basis for an undergraduate physics lab, and is highly suited to an ``at-home'' or online lab. It requires only a mobile phone for the experiment and a computer for data analysis. Students can gain an authentic experience of experimental physics, and can see the relevance of the basic equations of rotational dynamics to the behaviour of a favourite object -- the mobile phone.

\begin{acknowledgments}
The authors acknowledge the undergraduate students and postgraduate tutors who enthusiastically participated in an experiment which was the basis of this article, as part of our new PHYS1001 online course in 2020. 
\end{acknowledgments}

%\appendix

% If your manuscript is conditionally accepted, the editors will ask you to
% submit your editable LaTeX source file.  Before doing so, you should move
% all tables and figure captions to the end, as shown below.  Tables come 
% first, followed by figure captions (with figure inclusions commented-out).
% Figures should be submitted as separate files, collected with the
% LaTeX file into a single .zip archive.

%\newpage   % Start a new page for tables

\end{document}